\input harvmac
\input epsf

\noblackbox

\def\ie{{\it i.e.} }
\def\bfone{\relax{\rm 1\kern-.35em 1}}
\def\inbar{\vrule height1.5ex width.4pt depth0pt}

\def\IC{\relax\,\hbox{$\inbar\kern-.3em{\rm C}$}}
\def\ID{\relax{\rm I\kern-.18em D}}
\def\IF{\relax{\rm I\kern-.18em F}}
\def\IH{\relax{\rm I\kern-.18em H}}
\def\II{\relax{\rm I\kern-.17em I}}
\def\IN{\relax{\rm I\kern-.18em N}}
\def\IP{\relax{\rm I\kern-.18em P}}
\def\IQ{\relax\,\hbox{$\inbar\kern-.3em{\rm Q}$}}
\def\us#1{\underline{#1}}
\def\IR{\relax{\rm I\kern-.18em R}}
\font\cmss=cmss10 \font\cmsss=cmss10 at 7pt
\def\ZZ{\relax\ifmmode\mathchoice
{\hbox{\cmss Z\kern-.4em Z}}{\hbox{\cmss Z\kern-.4em Z}}
{\lower.9pt\hbox{\cmsss Z\kern-.4em Z}}
{\lower1.2pt\hbox{\cmsss Z\kern-.4em Z}}\else{\cmss Z\kern-.4em
Z}\fi}
\def\a{\alpha} \def\b{\beta}

\def\nup#1({Nucl.\ Phys.\ $\us {B#1}$\ (}
\def\plt#1({Phys.\ Lett.\ $\us  {B#1}$\ (}
\def\cmp#1({Comm.\ Math.\ Phys.\ $\us  {#1}$\ (}
\def\prp#1({Phys.\ Rep.\ $\us  {#1}$\ (}
\def\prl#1({Phys.\ Rev.\ Lett.\ $\us  {#1}$\ (}
\def\prv#1({Phys.\ Rev.\ $\us  {#1}$\ (}
\def\mpl#1({Mod.\ Phys.\ Let.\ $\us  {A#1}$\ (}
\def\ijmp#1({Int.\ J.\ Mod.\ Phys.\ $\us{A#1}$\ (}
\def\jag#1({Jour.\ Alg.\ Geom.\ $\us {#1}$\ (}
\def\tit#1|{{\it #1},\ }

\def\Coe#1.#2.{{#1\over #2}}
\def\coeff#1#2{\relax{\textstyle {#1 \over #2}}\displaystyle}
\def\coe#1.#2.{\relax{\textstyle {#1 \over #2}}\displaystyle}

\def\shalf{\relax{\textstyle {1 \over 2}}\displaystyle}

\def\Fh#1{{F\left(\shalf,\coeff{3}{4};\coeff{7}{4};#1\right)}}
\def\Gh#1{{F\left(\shalf,-\coeff{1}{4};\coeff{3}{4};#1\right)}}

\lref\thooft{G. 't Hooft, {\it A planar diagram theory for strong 
interactions}, Nucl. Phys. {\bf B72} (1974) 461.}
\lref\Polyakov{ A. Polyakov, {\it String theory and quark confinement},
hep-th/9711002.}
\lref\Malda{J. Maldacena, {\it The large N limit of superconformal
field theories and supergravity}, hep-th/9711200.}

\lref\wilson{ K. Wilson, Phys. Rep. {\bf 23} (1975) 331.}

\lref\camabi{C. Callan and J. Maldacena, {\it Brane dynamics from the
Born Infeld action}, C. Callan and J. Maldacena. 
hep-th/9708147.  }

\lref\ori{ O. Ganor,
{\it Six dimensional tensionless strings in the large N limit},
 Nucl. Phys. {\bf B489} (1997) 95,
 hep-th/9605201.}
\lref\IMSY{ N. Itzhaki, J. Maldacena, J. Sonnenschein and S. 
Yankielowicz, {\it Supergravity and the large N limit of 
theories with 16 supercharges}, hep-th/9802042. }
\lref\Wittenhol{E. Witten, {\it Anti de Sitter space and holography}, 
hep-th/9802150.}
\lref\GKP{ S. Gubser, I. Klebanov and A. Polyakov,
{\it Gauge theory correlators from noncritical string theory},
 hep-th/9802109.}


\lref\dthree{
I. Klebanov, {\it Worldvolume approach to absorption by nondilatonic 
branes}, Nucl. Phys. {\bf B499} (1997) 217;
S. Gubser, I. Klebanov and A. Tseytlin,
{\it String theory and classical absorption by three-branes},
Nucl. Phys. {\bf B499} (1997) 217,
hep-th/9703040;
 J. Maldacena and A. Strominger, 
{\it Universal low energy dynamics for rotating black holes},   
Phys. Rev. {\bf D56} (1997) 4975,
hep-th/9702015; 
S. Gubser and I. Klebanov, {\it Absorption by branes and Schwinger terms
in the world volume theory}, Phys. Lett. {\bf B413} (1997) 41. 
}

\lref\KS{
S. Kachru and E. Silverstein, {\it 4d conformal field theories and
strings on orbifolds}, hep-th/9802183.}

\lref\zerotwo{E. Witten, Proceedings of Strings 95, hep-th/9507121; 
 A. Strominger, {\it Open p-branes}, 
  Phys. Lett. B {\bf 383} (1996) 44,
hep-th/9512059;
N. Seiberg, {\it Non-trivial fixed points of the renormalization
group in six dimensions}, 
Phys. Lett. B {\bf 390} (1996) 169, hep-th/9609161;
N. Seiberg and E. Witten, {\it Comments on string dynamics in 
six dimensions},  Nucl. Phys. B{\bf 471} (1996) 121,
 hep-th/9603003.}

\lref\gradsh{I.S.~Gradshteyn and I.M.~Ryzhik, {\it Table of Integrals, 
Series, and Products}, Fifth Edition, A.~Jeffrey, ed. (Academic Press:
San Diego, 1994).}

\lref\HO{ G. Horowitz and H. Ooguri, {\it Spectrum of large N 
gauge theory from supergravity}, hep-th/9802116.}
\lref\Berk{M.~Berkooz, {\it A supergravity dual of a (1,0) field theory in 
six-dimensions,} hep-th/9802195.}
\lref\LNV{A.~Lawrence, N.~Nekrasov, and C.~Vafa, {\it On conformal field 
theories in  four-dimensions,} hep-th/9803015.}
\lref\CKKTV{P.~Claus, R.~Kallosh, J.~Kumar, P.~Townsend, and A.~V. Proeyen,
  {\it Conformal theory of {M2, D3, M5 and D1-branes + D5-branes},}
  hep-th/9801206.}

\lref\FFI{S.~Ferrara and C.~Fronsdal, {\it Conformal {M}axwell theory as a 
singleton field theory on {$AdS(5)$, IIB} three-branes and duality,} 
hep-th/9712239.}

\lref\FFII{
S.~Ferrara and C.~Fronsdal, {\it Gauge fields as composite boundary
  excitations,} hep-th/9802126.}
\lref\FFZ{
S.~Ferrara, C.~Fronsdal, and A.~Zaffaroni, {\it On {$N=8$ supergravity on
  $AdS(5)$ and $N=4$} superconformal yang-mills theory,} hep-th/9802203.}

\lref\FlF{
M.~Flato and C.~Fronsdal, {\it Interacting singletons,} hep-th/9803013.}

\lref\RYII{S.-J. Rey and J.~Yee, {\it Macroscopic strings as heavy quarks in 
large {N}
  gauge theory and anti-de Sitter supergravity,} hep-th/9803001.}

\lref\MaldaWL{J.~M.~Maldacena, {\it Wilson loops in large N field theories},
hep-th/9803002.}
\lref\AOY{O.~Aharony, Y.~Oz and Z.~ Yin, {\it 
M Theory on $AdS_p \times S^{11-p}$ and Superconformal Field Theories}, 
hep-th/9803051.}
\lref\Minw{S.~Minwalla, {\it Particles on $AdS_{4/7}$ and Primary Operators 
on $M_{2/5}$ Brane Worldvolumes}, hep-th/9803053.}
\lref\FZ{S. Ferrara and  A. Zaffaroni, {\it N=1,2 4D Superconformal Field 
Theories and Supergravity in $AdS_5$}, hep-th/9803060.}
\lref\DLP{M.J. Duff, H. Lu and C.N. Pope, 
{\it Gauge theory $\to$ IIB$\to$ IIA$\to$  M duality}, hep-th/9803061.}
\lref\LR{R.~Leigh and M.~ Rozali, {\it The Large $N$ Limit of the (2,0) 
Superconformal Field Theory}, hep-th/9803068.}
\lref\BKV{M.~Bershadsky, Z.~ Kakushadze and C.~Vafa, {\it String Expansion 
as Large N Expansion of Gauge Theories}, hep-th/9803076.}
\lref\Halyo{Edi Halyo, {\it Supergravity on $AdS_{4/7} \times S^{7/4}$ and 
M Branes}, hep-th/9803077.}
\lref\GR{G.~Horowitz and S.~ Ross, {\it Possible Resolution of Black Hole 
Singularities from Large N Gauge Theory}, hep-th/9803085.}
\lref\ASY{O.~Aharony, J.~Sonnenschein and S.~Yankielowicz, {\it
Interactions of strings and D-branes from M theory}, \nup{474} (1996) 309,
hep-th/9603009.}
\lref\Schwarz{J.~H.~ Schwarz, {\it Lectures on Superstring and M Theory 
Dualities}, {\it Nucl. Phys. Proc. Suppl.} {\bf 55B} (1997) 1, hep-th/9607201.}
\lref\DM{K.~Dasgupta and S.~ Mukhi, {\it BPS Nature of 3-String Junctions},
hep-th/9711094.}
\lref\Sen{A.~Sen, {\it String Network}, hep-th/9711130.}
\lref\RY{S-J.~ Rey and J-T.~Yee, {\it BPS Dynamics of Triple $(p,q)$ String 
Junction}, hep-th/9711202.}
\lref\KL{M.~ Krogh and S.~Lee, {\it String Network from M-theory}, 
hep-th/9712050.}
\lref\MO{Y.~Matsuo and K.~Okuyama, {it BPS Condition of String Junction 
from M theory}, hep-th/9712070.}
\lref\Oren{O. Bergman, {\it Three-Pronged Strings and $1/4 $BPS States in 
$N=4$ Super-Yang-Mills}, hep-th/9712211.}
\lref\GHZ{M.~R.~Gaberdiel, T.~Hauer and B.~Zwiebach, {\it 
Open string - string junction transitions}, hep-th/9801205.}
\lref\CT{C.G. Callan, L. Thorlacius, {\it Worldsheet Dynamics of String 
Junctions}, hep-th/9803097.}

%
%
\Title{\vbox{
\hbox{USC-98/004}
\hbox{\tt hep-th/9803111}
}}{\vbox{\centerline{\hbox{Quark-Monopole Potentials in Large $N$ Super Yang-
Mills}}
\vskip 8 pt
\centerline{ \hbox{}}}}
\centerline{Joseph A.~Minahan}
\bigskip
\centerline{{\it Physics Department, U.S.C., University Park,
Los Angeles, CA 90089-0484, USA}}
\medskip
\centerline{and}
\medskip
\centerline{\it Institute for Theoretical Physics, UCSB, Santa Barbara,
 CA 93106}
\bigskip

We compute the quark-monopole potential for ${\cal N}=4$ super Yang-Mills
in the large $N$ limit.  We find an attractive potential that falls off
as $1/L$ and is manifestly invariant under $g\to 1/g$.  The strength
of the potential is less than the quark-antiquark and
monopole-antimonopole potentials.

\vskip .3in

\Date{\sl {March, 1998}}

%
\parskip=4pt plus 15pt minus 1pt
\baselineskip=15pt plus 2pt minus 1pt
%
\newsec{Introduction}

There has been a  flurry of activity in large $N$ conformal
Yang-Mills theories
\refs{\Polyakov\Malda\FFI\CKKTV\IMSY\FFII\FFZ\GKP\HO\Wittenhol\KS\LNV\RYII
\MaldaWL\Berk\AOY\Minw\FlF\FZ\DLP\LR\BKV\Halyo\BKV{--}\GR\GR}.  
This was started by Maldacena's remarkable observation\Malda\ 
that computations in strongly coupled large $N$ Yang-Mills with $N=4$
supersymmetry can be mapped onto tree level computations in type IIB 
supergravity living on $AdS^5\times S^5$.  The radii of the 5-sphere
and anti-de Sitter space are equal and are given by 
$(4\pi g^2N)^{1/4}=(g^2_{YM}N)^{1/4}$. 
 Hence,
if $g^2_{YM}N>>1$ then the corresponding supergravity theory is weakly coupled
and so tree level supergravity computations should give accurate results
for strongly coupled Yang-Mills.  

This identification of $N=4$ Yang-Mills with IIB Supergravity has
led to nontrivial predictions for large $N$ Yang-Mills.  These are
the first predictions in strongly coupled Yang-Mills that go beyond
the level of the BPS states.

In particular, Rey and Yee\RYII\ and Maldacena\MaldaWL were able to compute 
the  coulomb potential between
a very massive quark and antiquark at strong coupling.  
The basic idea was to compute
the Nambu-Goto action in an anti-de Sitter background for a static string
configuration.  The result is  an  energy that diverges, but after subtraction
of the quark masses, one   is left with a finite attractive
potential that falls off as $1/L$, where $L$ is the distance between
the quarks on the D3-branes.

In this paper we will extend the result of \RYII\ and \MaldaWL\ 
to the case of a massive
quark and monopole.  We consider a brane configuration consiting of $N$ D3
branes at the origin and another D3 brane taken out to infinity.  This 
describes an ${\cal N}=4$ $U(N)\times U(1)$ gauge theory.  The BPS spectrum
contains heavy quarks and monopoles that transform under the fundamental of
$U(N)$ and are charged under the $U(1)$ gauge symmetry.  
The relevant string configuration is a {\bf Y} junction
\refs{\ASY\Schwarz\DM\Sen\RY\KL\MO\Oren\GHZ{--}\CT},
where one string coming out of the junction is attached to one of the $N$ 
 D3 branes at the origin.  The other
two strings are attached to the  D3 brane at infinity, but at a finite
distance $L$ from each other along the brane\foot{Rey and Yee  considered
using a junction configuration to describe baryons\RYII}.    After subtraction
of the monopole and quark mass, we are left with a finite attractive
potential between the particles.  We find an explicit function for
the attractive potential as a function of the coupling which is manifestly
$S$ dual under $g\to 1/g$. 

That the potential is attractive is not surprising.  A quark with
mass $\phi$ can bind with a monopole of mass $\phi/g$ to form a dyon
with mass $\phi\sqrt{1+1/g^2}$.  We should also expect that the 
attractive potential between a quark and a monopole is not as strong
as the potential between a quark and an antiquark,  since the latter
two objects can annihilate completely.  We will show
that this potential
is indeeed less than the $Q\overline Q$ potential.

In section 2 we review the calculation in \refs{\RYII,\MaldaWL}.  
In section 3 we extend this to the problem of a quark-monopole pair.  


\newsec{Quark-Antiquark potential}

Let us review the  calculation in \refs{\RYII,\MaldaWL}.  We will
follow closely the argument of \MaldaWL.
The metric for type IIB string theory in the presence of $N$ D3-branes
was computed by Horowitz and Strominger.  As we move down the throat,
we can approximate the metric by that for $AdS^5\times S^5$ with radii
$R$.  For the Nambu-Goto computation we will use the Euclidean version:
\eqn\metric{
ds^2 = \alpha' \left[
{ U^2 \over R^2} ( dt^2 + dx_i dx_i) + R^2 { dU^2 \over U^2 }
+ R^2 d\Omega_5^2
 \right]
}
where $R=(4\pi g N)^{1/4}$ and $g$ is the string coupling.  The world-sheet
action for the string is
\eqn\wsaction{
S~=~{1\over 2\pi\a'}\int d\tau d\sigma 
\sqrt{\det\left[G_{MN}\partial_\a X^M \partial_\b X^N\right]}
}
where $G_{MN}$ is the metric in \metric.   To find a static configuration
we set $\tau=t$ and $\sigma=x$, where $x$ is a direction along the D3-branes. 
 We assume that one of the D3-branes has been taken out to $U=\infty$ and that
the string configuration starts and ends on this brane.
Then the action simplifies to 
\eqn\simpleaction{
S~=~{1\over 2\pi}\int dt dx\sqrt{(\partial_x U)^2+U^4/R^4}
}
Since the configuration is static, the integration over $t$ leads to
a constant $T$.  $U(x)$ is minimized as a function of $x$ if
\eqn\umin{
{U^4\over \sqrt{(\partial_x U)^2+U^4/R^4}}={U_0}^2/R^2
}
where $U_0$ is a constant to be determined.
Hence we can write $x$ as a function of $U$
\eqn\xeq{
x={R^2\over U_0}\int_1^{U/U_0} {dy\over y^2\sqrt{y^4-1}}.
}
We can then find $U_0$ by setting $x$ to $x=L/2$ which corresponds to
moving half way along the string.  The middle of the string is then
at $U=U_0$.  Hence we find that
\eqn\unought{
L={2R^2\over U_0}\int_1^{\infty} {dy\over y^2\sqrt{y^4-1}}=
{R^2\over U_0}~{(2\pi)^{3/2}\over \Gamma(1/4)^2}.
}

The total energy is found by substituting \xeq\ with the condition
\unought\ into \simpleaction.  The resulting energy is infinite, because
the masses of the quarks have been included in the total energy.  To find
the quark mass, consider a string configuration that runs from
a D3 brane at large but finite $U=U_{max}$, to the $N$ D3 branes at $U=0$,
and with a fixed coordinate $x_i$ in the directions along the branes.  
Hence we can replace $\sigma$ by $U$ in \wsaction\ and we find that
the energy of this configuration is $U_{max}/2\pi$.  This is the quark mass.
Subtracting this off from the total energy and letting $U_{max}\to\infty$,
we find that the remaining energy is
\eqn\coulomb{
E_{Q\overline Q}~=~{2\over2\pi}
\int_{U_0}^\infty dU \left({U^2/{U_0}^2\over\sqrt{U^4/{U_0}^4-1}}-1\right)
~-~U_0~=~-U_0 {\sqrt{2\pi}\over(\Gamma(1/4))^2}=
~-~{4\pi^2({g^2_{YM}}N)^{1/2}\over L(\Gamma(1/4))^4}.
}

The result in \coulomb\ is consistent with $S$ duality.  To see this
consider the situation where the quarks are replaced with monopoles.
We then want to find a minimum energy configuration for this system.
Now instead of a fundamental string, we should consider a D-string attached
to the brane out at infinity.  The world sheet action is exactly the
same as in \wsaction, except for an extra factor of $1/g=4\pi/{g^2_{YM}}$.  
Hence the coulomb energy for the monopole-monopole pair is
\eqn\coulombMM{
E_{M\overline M}~=-~{16\pi^3({g^{-2}_{YM}}N)^{1/2}\over L(\Gamma(1/4))^4}.
}
Hence, under the $S$ dual transformation $g\to 1/g$ 
the potential in \coulomb\ is transformed into the potential in \coulombMM.  

Note that the results in \coulomb\ and \coulombMM\ are both valid if
$gN>>1$ and $N/g>>1$.  This is certainly true if $N$ is
large and $g\sim 1$.  

\newsec{The Quark-Monopole Potential}

In this section we derive the potential for a heavy quark and monopole.

As in the previous section, we have $N$ D3 branes at $U=0$
and one D3 brane at $U=U_{max}\to\infty$.  
We assume that there is a heavy quark
at $x=0$ and a heavy monopole at $x=L$.  Both the quark and monopole 
transform under the fundamental representation of $SU(N)$.
The string configuration
looks as follows:  In the $(x,U)$ plane we have a fundamental string
(\ie a string with $(p,q)$ charge $(1,0)$) attached to the D3 brane
at $(0,U_{max})$ and a D-string (\ie a $(0,1)$ string)  attached
to the D3 brane at $(L,U_{max})$.  These two strings are attached to
each other at the point $(\Delta L,U_0)$.  However, in order that the
$(p,q)$ charge is conserved, there must be another string with outgoing charge
$(1,1)$ attached to the other strings at the point $(\Delta L,U_0)$.
The other end of this string is attached to one of the D3 branes at 
$(\Delta L,U_0)$.  This configuration is shown in Figure 1.
\goodbreak\midinsert
\centerline{\epsfysize2in\epsfbox{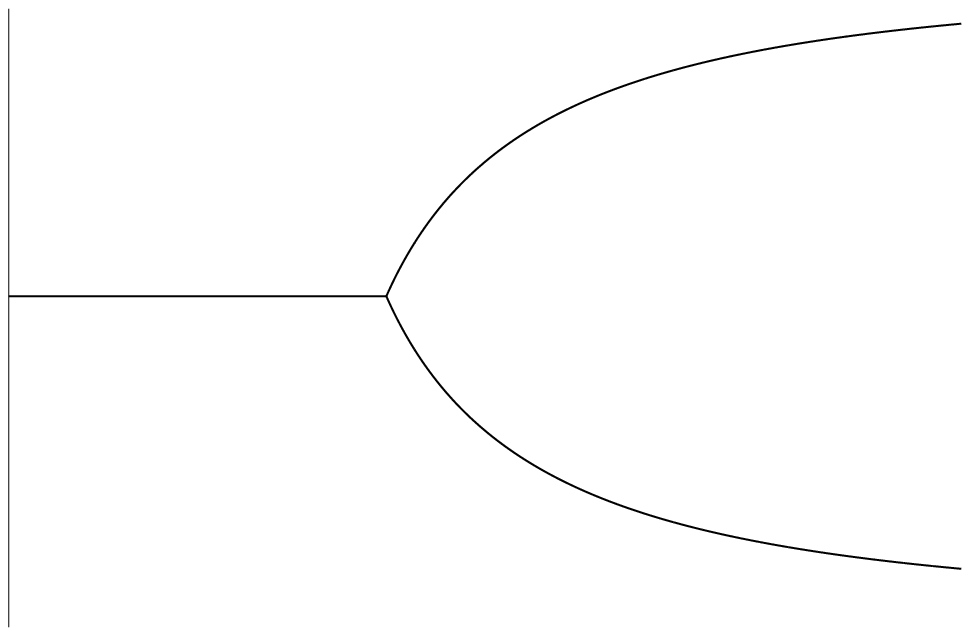}}
\leftskip 2pc
\rightskip 2pc\noindent{\ninepoint\sl \baselineskip=8pt {\bf Fig.~1}:
Three string-junction, with $(1,0)$ and $(0,1)$ strings coming in from
$U=\infty$ and the $(-1,-1)$ string attached to the $N$ D3 branes at $U=0$.}
\endinsert

{}From the Yang-Mills perspective on the branes, a quark and a monopole 
can 
bind together to form a dyon, so when the distance separation is zero,
we should be left with only with a $(1,1)$ string that stretches from
$U=0$ to $U=\infty$.   As
the quark and the monopole move  apart, the $(1,1)$ string
remains attached to one of the branes but its length gets shorter.
From the string picture, this sort of looks like
we are pulling open a zipper.

We can now turn to the world-sheet action to find a minimum energy 
configuration.  Since the dilaton field is constant for a background
with parallel D3 branes, the world sheet action for any $(p,q)$ string
in this background is given by \wsaction, multiplied by a factor
of $\sqrt{p^2+q^2 t^2}$, where $t=1/g$ and we are assuming that the
theta angle is zero.  Clearly, the $(1,1)$ string contributes
$U_0\sqrt{1+t^2}/(2\pi)$ to the total energy.
As for the other two strings, we need to minimize the action in
\simpleaction\ for each string, giving us the equations
\eqn\uminqm{
{U^4\over \sqrt{(\partial_x U)^2+U^4/R^4}}={U_i}^2/R^2\qquad\qquad
{i=1,2},
}
where $i=1$ ($i=2)$ is the result for the fundamental string (D-string).
Notice that $U_i$ is not necessarily equal to $U_0$.  Proceeding as before,
we find that the lengths of the two strings are
\eqn\lengths{\eqalign{
\Delta L&~=~{\a_1R^2\over U_0}\int_{\a_1}^\infty {dy\over y^2\sqrt{y^4-1}}\cr
L-\Delta L&~=~{\a_2R^2\over U_0}\int_{\a_2}^\infty {dy\over y^2\sqrt{y^4-1}}\cr
}}
where $\a_i=U_0/U_i$.  Notice that  $\a_i\ge 1$, otherwise the
integral will not be real. If $\a_i>1$, then the tangent vector
along the string will jump when going from the $(1,0)$ string to the
$(0,1)$ string.

We can now plug these expressions back into \wsaction\ and find the
energy for this configuration.  As in the quark-quark case, the energy
diverges because of the quark and monopole masses.  Subtracting off
$U_{max}/(2\pi)$ for the quark mass and $t U_{max}/(2\pi)$ for the
monopole mass, we are left with a finite energy, given by
\eqn\energyQM{
E_{QM}~=~ {U_0\over2\pi}\left[-1+
{1\over\a_1}\int_{\a_1}^\infty dy\left({y^2\over\sqrt{y^4-1}}
-1\right)-t+{t\over\a_2}\int_{\a_2}^\infty dy\left({y^2\over\sqrt{y^4-1}}
-1\right)+\sqrt{1+t^2}\right]
}
We can use \lengths\ to solve for $U_0$ in terms of $\a_1$ and $\a_2$.  
Substituting this into \energyQM\ we   can rewrite $E_{QM}$ as
\eqn\eQM{
E_{QM}~=~{R^2\over2\pi L}(F(\a_1)+F(\a_2))(G(\a_1)+tG(\a_2)+\sqrt{1+t^2}),
}
where the functions $F(\a)$ and $G(\a)$ can be written in terms of
the Gaussian hypergeometric functions and are given by
\eqn\expFG{\eqalign{
F(\a)&~=~{1\over3\a^2}\Fh{\coeff{1}{\a^4}}\qquad G(\a)=-\Gh{\coeff{1}{\a^4}}
}}

We now want to adjust $\a_1$ and $\a_2$ such that $E_{QM}$ is minimized.
We can take derivatives with respect to $\a_1$ and $\a_2$, set them to 
zero, and solve for $\a_1$ and $\a_2$.  In the end this seems to involve
proving an obscure identity for hypergeometric functions (see the appendix).

A much easier way to proceed is to adjust $\a_1$ and $\a_2$ such that
the net force  at the string junction is zero.  If this  were not
zero, then the junction could move and lower the energy.
For the $(1,0)$ and $(0,1)$ string, the derivatives $\partial_x U$ at $U=U_0$
are $\partial_x U=-{U_0}^2\sqrt{(U_0/U_1)^4-1}/R^2$ and 
$\partial_x U={U_0}^2\sqrt{(U_0/U_2)^4-1}/R^2$.
Moreover the infinitesmal lengths squared along the strings are
$ds^2=\a'{U_0}^6/({U_1}^4R^2)dx^2$ and $ds^2=\a'{U_0}^6/({U_2}^4R^2)dx^2$.
Hence, from \simpleaction\ we see that the tensions of the  strings
at $U=U_0$ are
\eqn\tension{
T_{1,0}={1\over2\pi}{{U_0}^4\over {U_1}^2R^2}{dx\over ds}~=~
{1\over2\pi\sqrt{\a'}R}U_0\qquad\qquad T_{0,1}={1\over2\pi\sqrt{\a'}R}tU_0
}
The tension of the $(1,1)$ string is 
$U_0(2\pi\sqrt{\a'}R)^{-1}\sqrt{1+t^2}$.  Therefore, the forces exerted
by each of the strings in the $x-U$ plane are
\eqn\forces{\eqalign{
\vec F_{1,0}&=\left(-(U_1/U_0)^2,\sqrt{1-(U_1/U_0)^4}\right)
{U_0\over2\pi\sqrt{\a'}R}\cr
\vec F_{0,1}&=\left((U_2/U_0)^2,\sqrt{1-(U_2/U_0)^4}\right)
{U_0~t\over2\pi\sqrt{\a'}R}\cr
\vec F_{-1,-1}&=(0,-1){U_0\sqrt{1+t^2}\over2\pi\sqrt{\a'}R}.
}}
Clearly the net force is zero if ${\a_1}^4=(U_0/U_1)^4=(1+t^2)/t^2$ and
${\a_2}^4=(U_0/U_2)^4=1+t^2$.
Substituting these expressions
back into \eQM, we find the expression
\eqn\coulombQM{\eqalign{
E_{QM}~&=~-{\sqrt{4\pi N}\over6\pi L\sqrt{g(1+g^2)}}
\left(g\Gh{\coeff{1}{1+g^2}}
+\Gh{\coeff{g^2}{1+g^2}}-\sqrt{1+g^2}\right)\cr
&\qquad\qquad\times\left(\Fh{\coeff{1}{1+g^2}}+g\Fh{\coeff{g^2}{1+g^2}}\right).
}}

As in the quark-quark case, the potential falls off as $1/L$, as is required by
conformal invariance.  Although perhaps not immediately obvious from the
expression, $E_{QM}$ is negative for all $g$ (A graph of $E_{QM}$ versus
$g$ is shown in Figure 2).  Moreover, $E_{QM}$ is manifestly
invariant under the $S$ duality transformation $g\to 1/g$, even though
strictly speaking, the binding energy is not invariant.  For large
but finite $U_{max}$, the binding energy is 
$B_{QM}=(\sqrt{1+1/g^2}-1-1/g)U_{max}/(2\pi)$,
so under $g\to 1/g$, $B_{QM}\to gB_{QM}$.  Of course, in our calculation
$B_{QM}\to \infty$, so the $S$ duality invariance for the quark-monopole
potential is valid in the region where $L>> \sqrt{gN}(U_{max})^{-1}$ and 
$L>>\sqrt{N/g}(U_{max})^{-1}$.
\goodbreak\midinsert
\centerline{\epsfysize2in\epsfbox{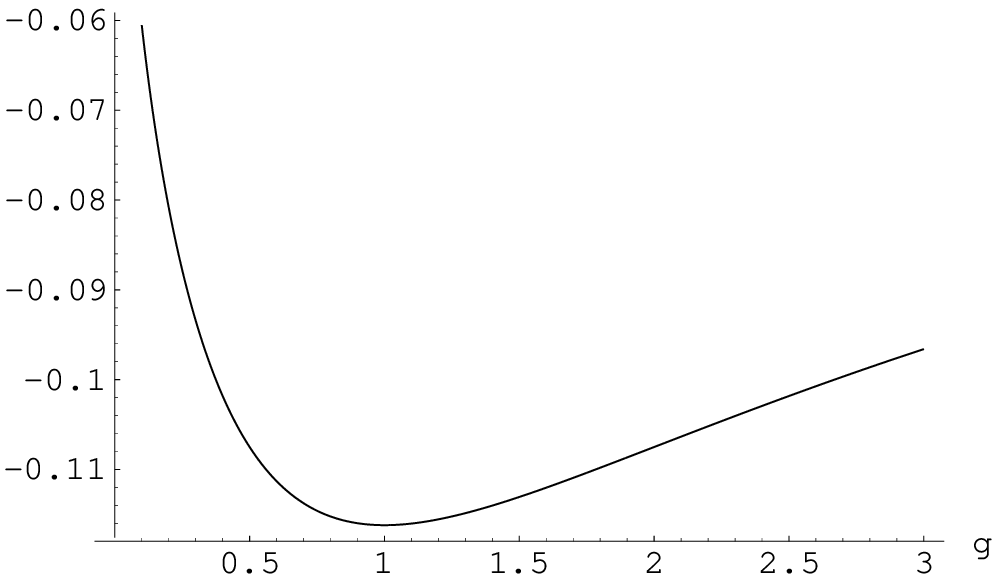}}
\leftskip 2pc
\rightskip 2pc\noindent{\ninepoint\sl \baselineskip=8pt {\bf Fig.~2}:
Plot of $E_{QM}L/\sqrt{N}$ vs. $g$.  $E_{QM}$ is negative for all $g$.}
\endinsert

Note that we can replace the $(0,1)$ and $(1,1)$
strings with the $(0,-1)$ and $(1,-1)$ strings.  This corresponds to
replacing the monopole with its charge conjugate.  The calculation proceeds
as before and we find the same attractive potential (at least when $\theta=0$).

It is instructive to compare the quark-monopole potential with the quark-quark
potential at the self dual point.  Letting $g=1$, one can easily evaluate
the expressions in \coulomb\ and \coulombQM, and find that 
$E_{QM}\approx .143 E_{Q\overline Q}$ at $g=1$.  This is consistent with
the reasoning presented above.

There is also interesting behavior as $g$ becomes large or small (but still
satisfying $4\pi gN>>1$ and $4\pi N/g>>1$).  As $g\to\infty$, we have
\eqn\asympFG{\eqalign{
\Fh{\coeff{g^2}{1+g^2}}&~=~{3\sqrt{2\pi^3}\over(\Gamma(1/4))^2}+{\rm O}(g^{-2})
\cr
\Gh{\coeff{1}{1+g^2}}&~=~1+{\rm O}(g^{-2})\cr
\Gh{\coeff{g^2}{1+g^2}}&~=~{\sqrt{2\pi^3}\over(\Gamma(1/4))^2}
+{\rm O}(g^{-2}).
  }}
Hence, we find that potential is approximately
\eqn\largeg{
E_{QM}~=~-{\pi^2(4\pi g^{-1}N)^{1/2}\over L(\Gamma(1/4))^4}+{\rm O}(g^{-3/2})
~=~-{4\pi^3({g^{-2}_{YM}}N)^{1/2}\over L(\Gamma(1/4))^4}+{\rm O}({g_{YM}}^{-3})
}
This is $1/4$ the potential for two monopoles.  Likewise, in the limit that
$g\to 0$, we find that the potential is $1/4$ the potential found for two
quarks.  


\goodbreak
\vskip2.cm\centerline{\bf Acknowledgements}
\noindent

I would like to thank J. Maldacena and other participants of Dualities in
String Theory '98 for  
valuable
discussions.  I would also like to thank the ITP at Santa Barbara for its
 hospitality during the course of this work.
This work was supported in part
by funds provided by the DOE under grant number DE-FG03-84ER-40168.

\goodbreak

\appendix{A}{Minimizing the energy}

Taking an $\a_1$ derivative on $E_{QM}$ in \eQM\ and \expFG, we
find
\eqn\eQMda{\eqalign{
{\partial E_{QM}\over \partial\a_1}&~=~-{R^2\over 6\pi L}
{1\over{\a_1}^3\sqrt{{\a_1}^4-1}}\Bigg[\left(\sqrt{t^2+1}-
\Gh{\coeff{1}{{\a_1}^4}}-
t\Gh{\coeff{1}{{\a_2}^4}}\right)\times\cr
&\qquad\qquad\times\left(3\a_1^2-
\sqrt{{\a_1}^4-1}~\Fh{\coeff{1}{{\a_1}^4}}\right)\cr
&\qquad\qquad+\left(1-\sqrt{{\a_1}^4-1}~\Gh{\coeff{1}{{\a_1}^4}}\right)\times
\cr
&\qquad\qquad\times
\left(\Fh{\coeff{1}{{\a_1}^4}}+{{\a_1}^2\over{\a_2}^2}\Fh{\coeff{1}{{\a_2}^4}}
\right)\Bigg].
}}
If we now let $\a_1=\sqrt{\sqrt{1+t^2}/t}$ and $\a_2=\sqrt{\sqrt{1+t^2}}$,
then \eQMda\ reduces to 
\eqn\eQMdb{\eqalign{
{\partial E_{QM}\over \partial\a_1}&~=~-{R^2\over 6\pi L}
\left({t^2\over1+t^2}\right)^{3/4}\times\cr
&\qquad\times\Bigg[
t\Fh{\coeff{t^2}{1+t^2}}\Gh{\coeff{1}{1+t^2}}-
t^{-1}\Fh{\coeff{1}{1+t^2}}\Gh{\coeff{t^2}{1+t^2}}\cr
&\qquad
+{\sqrt{1+t^2}\over t}\Fh{\coeff{1}{1+t^2}}-{\sqrt{1+t^2}}
\Big(\Gh{\coeff{t^2}{1+t^2}}\cr
&\qquad+t\Gh{\coeff{1}{1+t^2}}\Big)+3(1+t^2)\Bigg]
}}
The term inside the square brackets is identically zero.  A similar
calculation can be done for ${\partial E_{QM}\over \partial\a_1}$.

\listrefs
\end